\documentclass[aps,pra,epsfigure,twocolumn,showpacs]{revtex4}
\usepackage{dcolumn}    
\usepackage{bm} 
\usepackage{graphicx}
\usepackage{amsmath}    
\usepackage{latexsym}
\usepackage{amsfonts}   
\usepackage{amssymb}
\usepackage{array}      
\usepackage{epsfig}
\usepackage{epstopdf}

\newcommand{\ket}[1]{\left\vert#1\right\rangle}

\newcommand{\bra}[1]{\left\langle#1\right\vert}

\newcommand{\nbar}{\overline{n}}

\newcommand{\beq}{\begin{equation}}
\newcommand{\eeq}{\end{equation}}
\newcommand{\bea}{\begin{eqnarray}}
\newcommand{\eea}{\end{eqnarray}}

\begin{document}
\title{Non-Markovian effects on the non-locality of a qubit-oscillator system}
\author{Jie Li$^1$, Gerard McKeown$^1$, Fernando L. Semi\~ao$^2$, and Mauro Paternostro$^1$}
\affiliation{$^1$Centre for Theoretical Atomic, Molecular and Optical Physics, School of Mathematics and Physics, Queen's University, Belfast BT7 1NN, United Kingdom\\
$^2$ Centro de Ci\^encias Naturais e Humanas, Universidade Federal do ABC, 
R. Santa Ad\'elia 166, 09210-170 Santo Andr\'e, S\~ao Paulo, Brazil}

\begin{abstract}
Non-Markovian evolutions are responsible for a wide variety of physically 
interesting effects. Here, we study non-locality of the non-classical 
state of a system consisting of a qubit and an oscillator exposed to the effects of non-Markovian evolutions. We find that the different facets of non-Markovianity affect non-locality in different and non-obvious ways, ranging from pronounced insensitivity of the Bell function to quite a spectacular evidence of information kick-back.
\end{abstract}
\date{\today}
\maketitle

In 1935, E. Schr\"odinger formulated a thought experiment addressing some paradoxical implications of the Copenhagen interpretation of quantum mechanics when pushed to the realm of everyday experience~\cite{schro}. By describing a situation where the degrees of freedom of a `large' object are correlated in a quantum mechanical way to a `small' quantum system, the paradox by Schr\"odinger (commonly referred to as the `cat paradox') embodies a genuine example of the possibility to enforce quantum features beyond the microscopic domain. Notwithstanding its almost octogenarian history, the cat paradox still defies a full understanding of its implications~\cite{jacob}. 

The steady-pace experimental progresses in quantum control achieved in the last twenty years have been able to produce instances very close to the original formulation by Schr\"odinger and are expected to help significantly in the grasping of fundamental concepts such as the quantum-to-classical transition, as well as the development of quantum technological applications~\cite{wineland}. States having the form 
\begin{equation}
\label{state}
|\psi\rangle=(|{\uparrow,D}\rangle_{sO}+|{\downarrow,{-}D}\rangle_{sO})/{\sqrt{2}},
\end{equation}
where $\{\ket{\uparrow}_{s},\ket{\downarrow}_s\}$ are the energy eigenstates of a spin-$1/2$ particle (a qubit) and $\ket{\pm{D}}_O$ are opposite-phase coherent states of a harmonic oscillator~\cite{barnett}, are faithful instances of the situations envisaged in Refs.~\cite{schro} and have been demonstrated in trapped-ion settings~\cite{wineland,ulee}. They are accessible (or close to be such) in other experimental contexts involving the effective interaction between spin-like systems and mechanical oscillators~\cite{vacanti,nanooscillator} or the all-optical generation of micro-macro states~\cite{rome}. In the first instance, one would consider effective two-level systems (such as neutral or artificial atoms) embedded in cavities endowed with movable light mirrors (embodying the continuous variable (CV) subsystem). In the second one, the spin and CV parts are provided by different degrees of freedom of two distinct photonic information carriers. Both settings are able to engineer states having the form of Eq.~\eqref{state} and both allow for the reconstruction of the Wigner function of the CV subsystem. As it will be seen in the next Section, such ability is crucial to the assessment of the Bell test at the core of our investigation. Remarkably, the multifaceted interests in studying quantum superposition states analogous to Eq.~\eqref{state} extend up to the assessment of environment-induced dynamical effects and their implications for the settlement, manipulation and protection of general quantum correlations. This is even more relevant when non-trivial environmental influences of a non-Markovian nature, such as those due to lack of divisibility dynamics and/or memory-keeping and feedback-inducing system-environment mechanisms are considered~\cite{petruccione}. The working principles of such processes are still largely unexplored and are expected to be relevant in condensed matter set-ups involving artificial spins and mechanical modes. The experimental handiness of such states and the possibility to mimic the effects of non-trivial, memory-keeping environments in fully controllable linear-optics test-beds~\cite{paris}, make up for the possibility to acquire knowledge on the {\it true} behavior of the quantum features of state~\eqref{state}, when exposed to physical non-Markovian dynamics.

Motivated by these arguments, in this paper we address the influences that non-Markovian dynamics giving rise to non-divisible maps have on the non-local nature of Eq.~(\ref{state}) by studying two different configurations. First, we analyze the effects imparted by a spring-like coupling between the CV part of our state and an ensemble of quantum harmonic oscillators modelling quantum Brownian motion~\cite{Hu}. We then move to an effective post-Markovian dynamics of the spin-part only, as modelled by the master equation (ME) proposed in Ref.~\cite{SL} and analyzed, for a single-qubit problem in Ref.~\cite{Man}. 
Evident signatures of non-Markovianity have been found in the trend of entanglement and discord~\cite{zurekvedral} for an initially quantum-correlated state of two harmonic oscillators~\cite{Vasile}. Here, not only we analyze a different figure of merit and form of quantum correlations, but we also address a radically different clas of states. We find that the behavior of the Bell function associated with a Brownian motion-affected superposition state of a qubit and an oscillator shows quite subtle features. First, Brownian motion affects the non-local nature of such state in quite a significant way when the cut-off frequency of the Brownian bath is much smaller than the natural oscillation frequency of the CV subsystem, {\it i.e.} in the regime that would correspond to a strong non-Markovian limit: large-amplitude revival peaks are found, showing the kick-back mechanism that the memory-keeping environment can exert over the system. Second, the post-Markovian ME
turns out to be unable to induce a non-monotonic decay of the
Bell function. Yet, such dynamics is nondivisible,
as it is straightforward to check, and as such it deviates
from the prescriptions commonly accepted for Markovianity.
Indeed, it is revealed as fully non-Markovian by the
measure recently proposed by Rivas et al.~\cite{rivas}. Remarkably,
our study provides indirect evidence that the evolution of nonlocality in a quantum superposition state of a qubit and an oscillator is qualitatively similar to what would arise from the
measure proposed by Breuer et al.~\cite{breuer}, which is designed to
point towards the back-flow of information from the environment
to the system.

The remainder of this manuscript is organized as follows: In Sec.~\ref{uno} we describe the formal tools that will be used in the core part of our analysis and discuss, very briefly, the case of Markovian evolutions. This will be used as a milestone for the comparisons with the explicitly non-Markovian cases. In Sec.~\ref{due}, of the other hand, we present the key part of our study and address non-locality in a quantum superposition state of a qubit and an oscillator under non-Markovian dynamical conditions. Finally, Sec.~\ref{conc} is for our conclusions and outlook. We delegate the most technical parts of our work to two appendices.

\section{Tools and Markovian benchmarks}
\label{uno}

We start our study by describing the formal approach to non-locality that will be used throughout this work, which is similar to the one proposed in Ref.~\cite{wodkiewicz} and used by Spagnolo {\it et al.} in Ref.~\cite{rome}. In our thought experiment, the spin of the discrete-variable component of the system is probed along a direction ${\bm n}{=}(\sin\theta,0,\cos\theta)$ of the Bloch sphere by the bi-dimensional operator  
\begin{equation}
 \hat{\sigma}(\theta)=\sin\theta\,\hat{\sigma}_x+\cos\theta\,\hat{\sigma}_z
 \end{equation}
with $\hat\sigma_{j}~(j{=}x,y,z)$ the $j$-Pauli operator. As discussed in Ref.~\cite{banaszek}, non-locality of the state of a CV system can be tested in the phase-space by using the dichotomic parity operator 
\begin{equation}
\hat{\Pi}=(-1)^{\hat{n}}=\sum^\infty_{n{=}0}(\ket{2n}\bra{2n}-\ket{2n+1}\bra{2n+1})
\end{equation}
 acted upon by the displacement $\hat {\cal D}(\beta)=\exp[\beta\hat a^\dag-\beta^* \hat a]$ ($\beta\in{\mathbb C}$)~\cite{barnett} so as to form the displaced parity operator $\hat{\Pi}(\beta)=\hat{\cal D}(\beta)\hat{\Pi}\hat{\cal D}^{\dagger}(\beta)$. Here, $\hat{n}=\hat a^\dag\hat a$ is the bosonic operator number of the harmonic oscillator whose annihilation (creation) operators is $\hat a$ ($\hat a^\dag$) and $\ket{n}$ is a Fock state with $n$ excitations. The key point of such phase-space approach is that $\langle\hat\Pi(\beta)\rangle=(\pi/{2})W(\beta)$, where $W(\beta)$ is the Wigner function associated with the state over which the expectation value of $\hat\Pi(\beta)$ is calculated. Therefore, we can easily construct the correlation function ${\cal C}(\beta,\theta){=}\bra{\psi}\hat\sigma({\theta}){\otimes}\hat\Pi(\beta)\ket{\psi}$ from which we get the Bell-Clauser-Horne-Shimony-Holt (CHSH) function 
\begin{equation}
{\cal B}(\beta,\theta;\beta',\theta'){=}{\cal C}(\beta',\theta'){+}{\cal C}(\beta',\theta){+}{\cal C}(\beta,\theta'){-}{\cal C}(\beta,\theta).
\end{equation}
Local realistic theories impose the bound $|{\cal B}|{\le}{2}$, which is violated by quantum mechanics, although not maximally, when state $\ket{\psi}$ is used and a judicious choice of the set of parameters $\{\beta,\theta;\beta',\theta'\}$ is made~\cite{wodkiewicz}. This form of the Bell-CHSH test has been shown to be effective in revealing the non-local nature of correlations in a state such as $\ket{\psi}$~\cite{wodkiewicz} (see also Spagnolo {\it et al.} in Ref.~\cite{rome}) and we thus believe it is a very appropriate tool for the understanding of the role that the interaction with an environment has on the non-local properties of such state. 

First, we set a benchmark by briefly addressing the case of a Markovian dynamics as encompassed by general amplitude damping (AD) and phase damping (PD) processes, both for the spin and the CV subsystem~\cite{nielsenchuang}. Regardless of the dimensionality of the system at hand, both cases are most effectively tackled by means of the operator-sum representation of a quantum channel. We call $\{\hat{\cal A}^p_{k}(t)\}$ the set of Kraus operators specifying the non-unitary process $p{=}{\rm AD,PD}$, so that the evolved state at time $t$ takes the general form $\rho(t){=}\sum_{k}\hat{\cal A}^p_{k}(t)\ket{\psi}\bra{\psi}\hat{\cal A}_{k}^{p\dagger}(t)$. The explicit Kraus operators for AD and PD and the form of the correlation functions corresponding to the case where the spin or CV components are influenced by the environment are provided in {Appendix~A}. Here, it is enough to study the behavior of the numerically optimized Bell function $\max_{\{\theta,\beta;\theta',\beta'\}}|{\cal B}|$ for the superposition state, which is shown in Fig.~\ref{ADPD} against the probability ${\cal P}_p$ of occurrence of the channel (introduced in Appendix~A). While, expectedly, affecting the CV with an AD channel results in a quicker decay of the Bell function as compared to the spin-affected scenario, phase damping seems to be oblivious to the dimensionality of the subsystem being influenced by the environment. This is likely to be due to the fact that, as $D$ gets sufficiently large, $\langle-D|D\rangle\simeq{0}$ and the CV state is effectively encoded in the fictitious quasi-qubit $\{\ket{\pm D}\}$ and the PD channel in the spin and CV-affected scenario give very similar outcomes. Needless to say, as the AD mechanism depends critically on the number of excitations in the parties entering a given state, such 'homogenization' does not take place in Fig.~\ref{ADPD} {\bf (a)}. In both cases, however, a monotonic decay of $\max|{\cal B}|$  is observed (for easiness of notation we omit the variables over which the optimization is performed).

\begin{figure}
{\bf (a)}\hskip4cm{\bf (b)}
\includegraphics[width=0.47\linewidth]{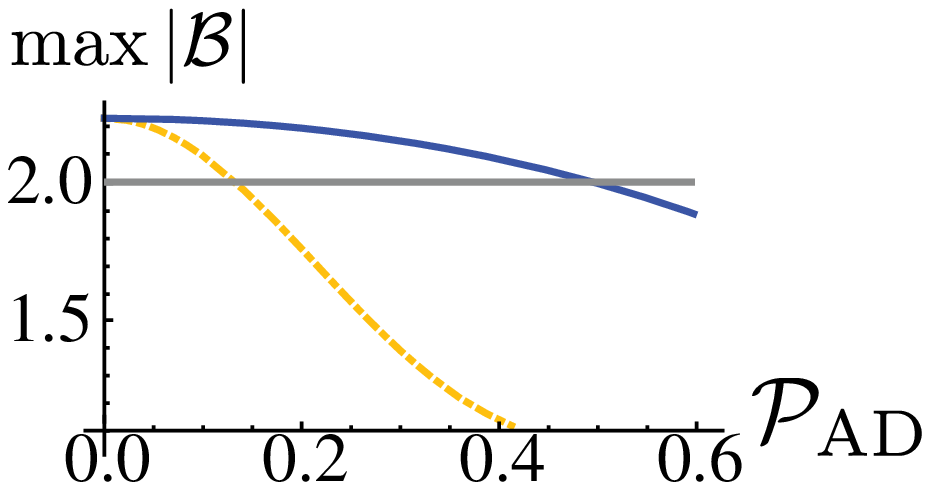}~~~\includegraphics[width=0.49\linewidth]{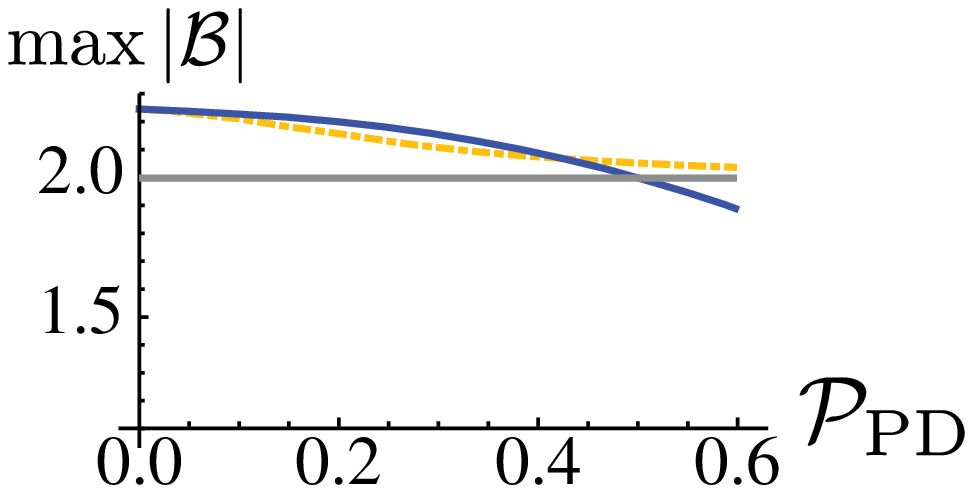}
\caption{We study $\max|{\cal B}|$ under the effects of amplitude- and phase-damping mechanisms. In {\bf (a)} [{\bf (b)}] $\max|{\cal B}|$ is affected by AD [PD] and studied against the probability ${\cal P}_{\rm AD}$ (${\cal P}_{\rm PD}$) for the environmental action to occur. In both panels, the solid curve (dot-dashed curve) is for the spin-affected (CV-affected) case. The horizontal line shows the local realistic bound and we have taken $D=2$.}
\label{ADPD}
\end{figure}

\section{Non-Markovian effects on non-locality}
\label{due}


Our aim is now  to study non-Markovian dynamics, looking for evidence of information kick-back over the evolution of the hybrid system at hand.  We start by analyzing the CV-affected case and consider the harmonic oscillator (with frequency $\omega_O$) as coupled to an $N$-mode bosonic environment according to the interaction Hamiltonian $\hat H_{{\rm CV}}{=}\hbar g\sum^{N}_{j=1}c_j\hat{q}\otimes\hat Q_j$, where $c_j$ is the coupling rate between the $j^{\rm th}$ mode and the CV subsystem and $g$ is a dimensionless coupling strength and $\hat q$ ($\hat Q_j$) is the position-like quadrature of the harmonic oscillator (the $j^{\rm th}$ mode of the environment). This model is known to give rise to quantum Brownian motion~\cite{Hu}. The effective evolution of the CV subsystem is thus regulated by the time-local ME 
\begin{equation}
\partial_t\rho_{O}{=-}(i/\hbar)\omega_O[\hat a^\dag\hat a,\rho_{O}]{+}{\cal L}_{bm}(\rho_{O}),
\end{equation}
where $\rho_{O}$ is the density matrix of subsystem $O$ and
\begin{equation}
\label{HPZ}
\begin{aligned}
{\cal L}_{bm}(\cdot){=}&{-}\Delta(t)[\hat q,[\hat q,\cdot]]{+}\Xi(t)[\hat q,[\hat p,\cdot]]\\
&{-}i\gamma(t)[\hat q,\{\hat p,\cdot\}]{+}\frac{ir(t)}{2}[\hat q^2,\cdot]
\end{aligned}
\end{equation}
that accounts for diffusion [at rates $\Delta(t)$ and $\Xi(t)$], damping [at rate $\gamma(t)$] and the renormalization of the frequency of the CV subsystem. We have introduced the bosonic momentum-like quadrature $\hat p$ of the $O$ subsystem. The derivation of Eq.~\eqref{HPZ} does not require the rotating-wave approximation nor it invokes the Born-Markov one. Various dynamical phases have been identified for entanglement and quantum discord of a two-mode Gaussian states under such dynamics~\cite{Vasile}. These include effects of entanglement sudden-death and revival, depending on the spectral/memory properties of the environment itself.

\begin{figure*}[t!]
{\bf (a)}\hskip7cm{\bf (b)}
\includegraphics[width=7cm]{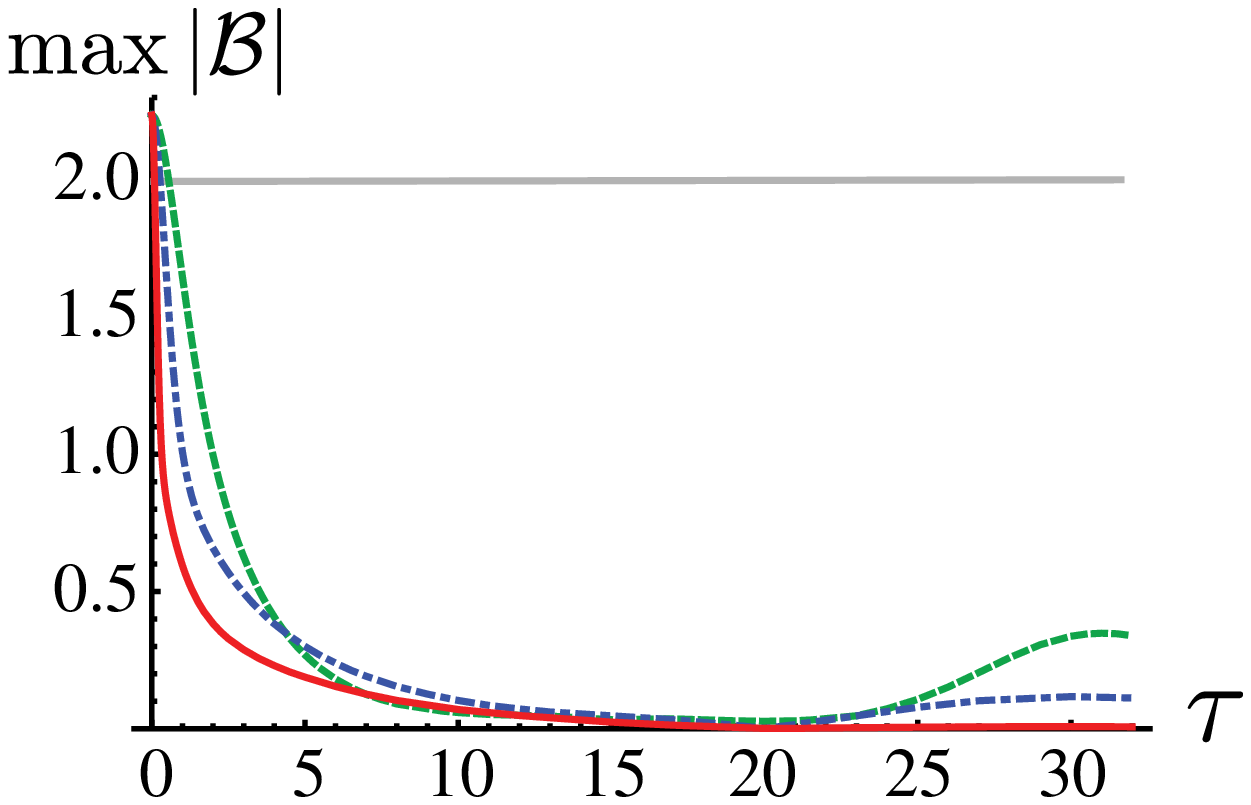}~~\includegraphics[width=7cm]{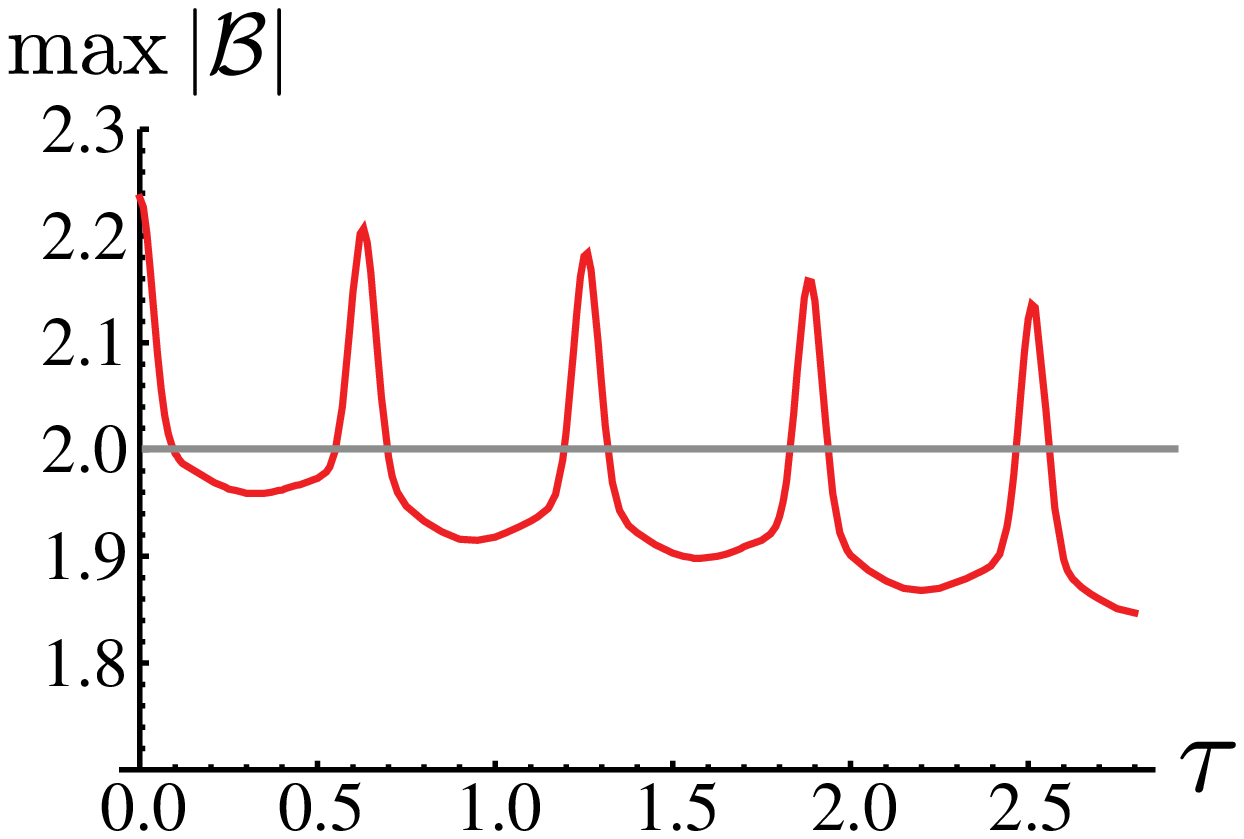}
\caption{Numerically optimized $|{\cal B}|$ plotted against $\tau$ for $D=2$, $k_B{\cal T}/\hbar\omega_c=25$. In panel {\bf (a)} we have taken $x=10$ and $g=0.3$ (solid line), 0.1 (dot-dashed line) and 0.05 (dashed one). Panel {\bf (b)}: $\max|{\cal B}|$ for $x=0.2$ and  $g=0.05$. The straight line shows the bound imposed by local realistic theories.}
\label{MKfigure}
\end{figure*}

Here, we focus on the case of weak coupling, Ohmic environmental spectral density with a cutoff $\omega_c$, high temperatures and short time-scale limit. Under these conditions, the contribution of $r(t)$ to the solution of the ME is negligible, and the elements of the vector of coefficients ${\bm v}(t)=(\Delta(t)\,\Xi(t)\,\gamma(t))^T$ entering ${\cal L}_{bm}$ are given by the expressions~\cite{Vasile}
\begin{equation}
{\bm v}_j(t){=}\frac{g^2\omega_Ox^2}{2(1+x^2)}f_j({\cal T})\left\{a_j{-}e^{-\tau}\!\!\left[a_j\cos\left(\frac{\tau}{x}\right){+}b_j\sin\left(\frac{\tau}{x}\right)\right]\right\}
\end{equation}
with $\tau{=}\omega_c t$, $x{=}\omega_c/\omega_O$, ${\cal T}$ the environmental temperature, ${\bm f}({\cal T}){=}({k_B{\cal T}}/{\hbar\omega_c})(1\,1\,\,{\hbar\omega_c}/k_B{\cal T})^T$ [$k_B$ is the Boltzmann constant], ${\bm a}{=}(x\,1\,1)^T$ and ${\bm b}{=}(-1\,x\,x)^T$. The Brownian ME can be exactly solved using a phase space approach~\cite{intravaia} that allows for the determination of the Wigner function 
$W(\beta)={\cal F}[\chi_t(\hat{\bm z})]$,
where ${\cal F}[\cdot]$ indicates the complex Fourier transform, $\hat{\bm z}{=}(\hat q\, \hat p)^T$ is the vector of quadratures and $\chi_t(\hat{\bm z})$ is the Weyl function associated with the initial CV state~\cite{barnett}. Explicitly
\begin{equation}
\chi_t(\hat{\bm z})=e^{-\hat{\bm z}^T\bar{W}(t)\hat{\bm z}}\chi_0(e^{-\Gamma(t)/2}R^{-1}(t)\hat{\bm z})
\end{equation}
with $\chi_0(\hat{\bm z}){=}{\rm Tr}[e^{i(\hat p\hat q-\hat q\hat p)}\rho_{O}(0)]$ the Weyl function of the CV state at $t{=}0$, $\Gamma(t){=}2\int_0^t\gamma(s)ds$, $R(t){=}\cos(\omega_Ot)\openone{+}i\sin(\omega_Ot)\hat\sigma_y$ and 
\begin{equation}
\bar{W}(t){=}\frac{e^{-\Gamma(t)}}{2}R(t)\!\!\left[\int_0^t\!\!\!\!e^{-\Gamma(s)}R^T(s)
{\cal M}(s)
R(s)ds\right] R^T(t),
\end{equation}
with ${\cal M}(s)=\left[\begin{matrix}
2\Delta(s)&\!\!-\Xi(s)\\
-\Xi(s)\!\!& 0
\end{matrix}\right]$. 
Due to the weak coupling and high temperature assumptions, $g$ should take small values, while $k_B {\cal T}/\hbar\omega_c$ cannot be too small. For short time scales, we can set $e^{\pm\Gamma(t)}{\simeq}1$~\cite{Vasile}.

With this at hand, we determine the Wigner functions $W^{ij}_t({\beta})$ of the CV components of the density matrix associated to the spin part $\rho^{ij}_t$ ($i,j{=}{\uparrow},{\downarrow}$).
  Fig.~\ref{MKfigure} shows $\max|{\cal B}|$ as a function of $\tau$ in two different dynamical regimes. For a large number of environmental mode-frequencies ({\it i.e.} for a large cutoff $\omega_c$) and a relatively large coupling strength, the numerically optimized Bell function shows a monotonically decreasing behavior that makes us lose evidences of non-local character of the superposition state very soon in time. Although by lowering $g$ we observe some very partial revival of $\max|{\cal B}|$ due to a kick-back of coherence into the spin-CV system, this is not sufficient to give rise to violate the local-realistic bound again. The trend changes dramatically for $\omega_c\ll\omega_0$ and $g\ll{1}$, which bring us to the phase of ``non-Markovian revivals'': $\max|{\cal B}|$ becomes a periodic function of time and shows slowly-fading peaks at which the hybrid Bell-CHSH inequality studied here is quite largely violated. Such oscillations, which are typical of non-Markovian dynamics, are related to the appearance of temporal regions where $\Delta(t)$ achieves negative values and are connected to the memory of the environmental system, which keeps track of the system's state and feeds this information back to it. 

\begin{figure*}[ht!]
{\bf (a)}\hskip7cm{\bf (b)}\\
\includegraphics[width=7cm]{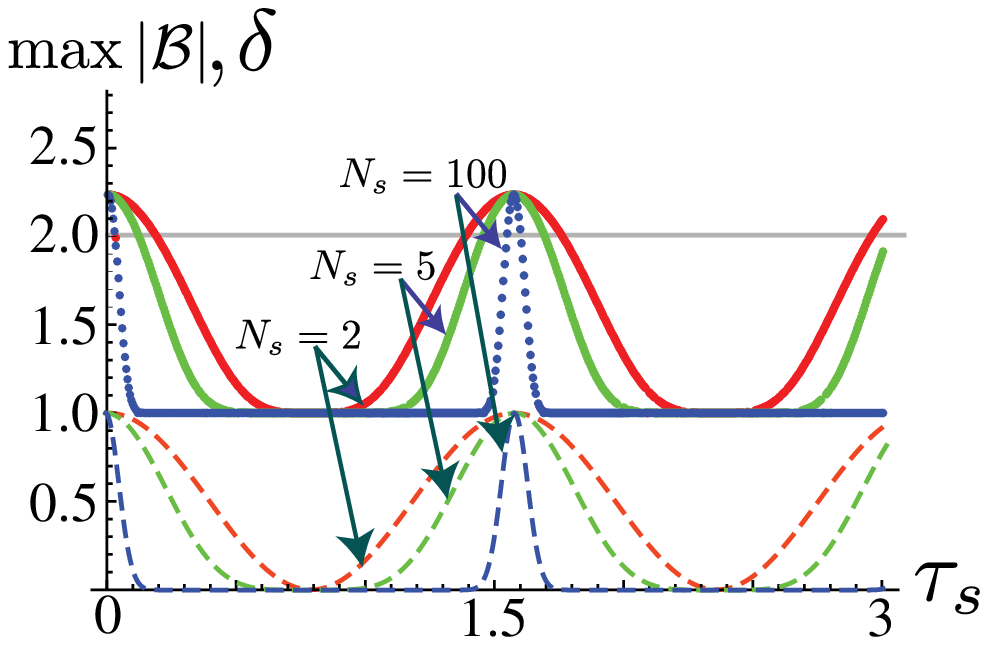}\includegraphics[width=7cm]{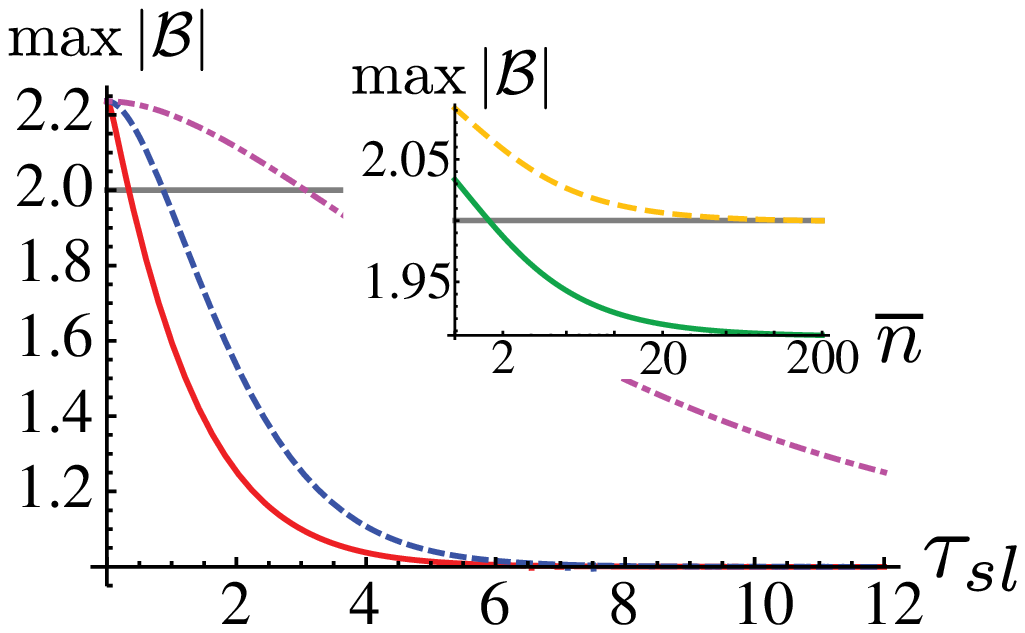}
\caption{{\bf (a)} Upper part: Time dependence of the numerically optimized Bell function for a quantum superposition state between a qubit and an oscillator with $D{=}2$ and whose $s$ subsystem is affected by  $N_s{=}2,5,100$ non-interacting spins according to the model $\hat{H}_{spin}$. Here, $\tau_s{=}At$ is a dimensionless interaction time. Lower part: Behavior of the trace distance $\delta(\tau){=}{\rm Tr}|\rho_{s,+}{-}\rho_{s,-}|/2$ (with $\rho_{s,\pm}{=}\ket{\pm}_s\!\bra{\pm}$ and $\hat\sigma_{x}\ket{\pm}_s{=\pm}\ket{\pm}_s$) upon which the measure of non-Markovianity proposed in~\cite{breuer} is built. From top to bottom curve, we have $N_s=2,5,100$. {\bf (b)} We show $\max|{\cal B}|$ against $\tau_{sl}$ for $\gamma_0/\gamma{=}0.05,1,10$ (solid, dashed and dot-dashed line, respectively) and $\nbar=0$. The inset shows $\max|{\cal B}|$ against $\nbar$ for $\tau_{sl}=1.6$ and $\gamma_0/\gamma=10,14.3$ (solid and dashed line, respectively). The horizontal line shows the local realistic bound. We have $D{=}2$ in all plots.}
\label{SpinBath}
\end{figure*}

The connection between non-Markovian revivals of non-locality and the memory properties of the environmental system appears to be reinforced by the study of a simple model for system-environment interaction. We now consider the spin-part $s$ of the superposition state as coupled to a star-like collection of $N_s$ non-interacting spin-$1/2$ particles via the energy-preserving longitudinal Ising model $\hat{H}_{spin}{=}\hbar A\sum^{N_s}_{k{=}1}\hat\sigma_z\otimes\hat\sigma_{z,k}$ [$\hat\sigma_{z,k}$ is the $z$-Pauli operator of the $k^{\rm th}$ environmental spin and $A$ is the corresponding coupling strength]. The evolution induced by $\hat H_{spin}$ over $s$ is exactly solvable due to its excitation-preserving, non-interacting nature. We assume the spin star as prepared in the maximally mixed state $\openone/2^{N_s}$, which leads to the reduced dynamics of spin $s$~\cite{breuer2}
\begin{equation}
\begin{aligned}
\rho_s(t)&=\rho^{\uparrow\uparrow}_s(0)\ket{\uparrow}_s\!\bra{\uparrow}+\rho^{\downarrow\downarrow}_s(0)\ket{\downarrow}_s\!\bra{\downarrow}\\
&+[\cos(2\tau_s)]^{N_s}(\rho^{\uparrow\downarrow}_s(0)\ket{\uparrow}_s\!\bra{\downarrow}+h.c.)
\end{aligned}
\end{equation}
with $\tau_s=At$, where $\rho^{ij}_s(0)$ ($i,j{=}{\uparrow},{\downarrow}$) are elements of density matrix $\rho_s(0)$. The calculation of the spin-oscillator correlation function proceeds now along the lines shown above and leads to the analytic form
\begin{equation}
\begin{aligned}
{\cal C}(\theta,\beta)&{=}e^{-2|\beta|^2}\{\sin\theta\cos(4D\beta_i) [\cos (2 \tau_s )]^{N_s}\\
&{+}e^{-2D^2}\cos\theta\sinh(4D\beta_r)\},
   \end{aligned}
\end{equation}
where $\beta_i{=}{\rm Im}[\beta]$, and $\beta_r{=}{\rm Re}[\beta]$. The oscillatory function of time appearing in ${\cal C}(\theta,\beta)$ immediately reveals the non-monotonic behavior that the corresponding Bell-CHSH function will exhibit, thus signalling non-Markovianity. Clearly, at $\tau_s{=}r\pi/2~(r{\in}{\mathbb Z})$, the decoherence factor $[\cos (2 \tau_s )]^{N_s}$ becomes ineffective, regardless of the number of environmental spins, and the correlation function achieves the value corresponding to a pure qubit-oscillator superposition state. Therefore, the non-Markovian non-locality revivals are full and the only effect of a growing size of the star is the narrowing of the revival peaks~\cite{commentosize}. This results in shorter time windows where the Bell-CHSH inequality is violated. This analysis is displayed in the upper part of Fig.~\ref{SpinBath} {\bf (a)}, where we assess the case of three different values of $N_s$.

An interesting remark is due, now, in relation to the aims of our study. This spin model for decoherence is known to provide a divergent value of the non-Markovian measure proposed by Breuer {\it et al.} in Ref.~\cite{breuer} [which is unbounded]. This means that the dynamics experienced by $s$ can never be described by a Markovian model. The model, in fact, is such that the trace distance of equatorially antipodal states of $s$ (upon which the measure in~\cite{breuer} is built) is given exactly by $|\cos (2 \tau_s )|^{N_s}$. This is clearly shown in the lower part of Fig.~\ref{SpinBath} {\bf (a)}. Therefore, not only the non-Markovian revivals of non-locality persist in time due to the infinitely non-Markovian nature of the evolution, but their occurrence is clearly related to the changes in the trace distance, thus providing a clear connection between the kick-back of information from the star system to $s$ and the revived non-local features of $\ket{\psi}$.

However, one should be careful in dealing with the relationship between non-Markovianity and non-locality, due to the multifaceted nature of the former statistical phenomenon: it would be a mistake to identify an in-principle non-Markovian dynamics with  the occurrence of non-monotonic trends in $\max|{\cal B}|$. We illustrate such point considering an important instance of post-Markovian dynamics proposed by Shabani and Lidar in~\cite{SL}. The model includes explicitly a memory 
kernel $k(t)$ that renders the ME time non-local as
\begin{equation}
\label{sleq}
\partial_t\rho_s(t)=\hat L\int_0^tk(t')e^{\hat Lt'}\rho_s(t-t')dt',
\end{equation}
where $\hat L[\cdot]$ is the standard Markovian Liouvillian describing dissipation at rate $\gamma_0$ induced by a thermal bath with mean occupation number $\nbar$~\cite{gardiner,dissipatore}. We focus on the widely used form of the memory kernel $k(t){=}\gamma e^{-\gamma t}$, a choice that guarantees complete positivity of the map for any value of $(\gamma,\gamma_0,\nbar)$ and at any instant of time~\cite{SL,Man2}. The dynamics in Eq.~\eqref{sleq} can be straightforwardly solved using a rather technical approach discussed in Refs.~\cite{Briegel,Man}. For the sake of providing a self-contined presentation, we summarize the steps needed for such a derivation in Appendix B. We eventually get an analytic form of the density matrix of the evolved superposition state and, thus, the correlation function, whose expression is however too cumbersome to be reported here. Fig.~\ref{SpinBath} {\bf (b)} shows the results of our quantitative study: we display the optimized Bell-CHSH function against the dimensionless interaction time $\tau_{sl}{=}\gamma_0t$ and for a long/short environmental memory time as compared to the dissipation time $\gamma^{-1}_0$. When $\gamma{\ll}\gamma_0$, the environment has a long-time memory, thus pushing the dynamics away from Markovianity. Interestingly, this  results in larger temporal windows where the violation of the Bell inequality is observed. Viceversa, for a larger $\gamma$, such window becomes shorter and the dynamics gets closer to a Markovian one. The resilience induced by a memory-keeping evolution turns out to be quite spectacular when studied against the thermal nature of the environment. As shown in the inset of Fig.~\ref{SpinBath} {\bf (b)}, while for $\gamma_0/\gamma{=}10$ it is enough to have $\nbar{=}1.6341$ to stop violating the Bell-CHSH inequality, a small raise to $\gamma_0/\gamma{\simeq}14.3$ is enough to push the value of $\nbar$ at which $\max|{\cal B}|{\le}2$ to $\nbar{\simeq}200$. 

It is remarkable, though, that besides such induced robustness, no otherwise evident signature of non-Markovianity (such as the ripples highlighted previously and typically exhibited by other indicators of quantumness~\cite{Vasile}) can be deduced from the analysis of the correlation function ${\cal C}(\theta,\beta)$ associated with this case and of Fig.~\ref{SpinBath} {\bf (b)}. We believe that the reason for such a behavior, though, should be researched in the multifaceted nature of non-Markovian dynamics.
In fact, the Shabani-Lidar ME can be recast into the form of a time-dependent, time-local ME whose coefficients are not all positive, therefore signaling the break-down of divisibility~\cite{mazzola} and thus the impossibility to describe the evolution in Markovian terms, according to the criterion put forward by Rivas {\it et al.}~\cite{rivas}, although no kick-back of information is possible (as witnessed by the fact that the measure proposed in~\cite{breuer} is strictly null, for such a dynamical map). It is just intriguing that the Bell-CHSH function is able to {\it experience} the subtleties of such differences in such a striking way.

\noindent
\section{Conclusions}
\label{conc}

We have studied the behavior of the Bell-CHSH function for a spin-oscillator superposition state under the influences of two models for non-Markovian dynamics, demonstrating the emergence of interesting features related to the various facets with which non-Markovianity manifests itself. Care should be used in assessing them by means of diverse non-classicality indicators, as they may turn up to give mutually inconsistent evidences. We have shown that the Bell-CHSH function is particularly sensitive to the subtleties of a dynamics that, although does not appear to be characterized by a kick-back of information from the environment, is nevertheless not divisible and as such far from being Markovian. It will be interesting to address similar questions in models for open-system dynamics specific of condensed matter set-ups, such as the $1/f$ noise that is expected to provide strong non-Markovian dynamical features. This will help us ascertain if the range of non-Markovian manifestations in non-locality tests is even richer than the predictions coming from our work.  

\noindent
\acknowledgments 

We thank L. Mazzola for invaluable advice. FLS thanks the Centre for Theoretical Atomic, Molecular and Optical Physics, Queen's University Belfast for the kind hospitality during completion of this work. MP thanks M.G.A. Paris for discussions. We acknowledge financial support from DEL and the UK EPSRC (EP/G004759/1). FLS  acknowledges  partial support from CNPq (Grant  No. 303042/2008-7) and the Brazilian National Institute of Science and Technology of Quantum Information (INCT-IQ).

\renewcommand{\theequation}{A-\arabic{equation}}
\setcounter{equation}{0}
\section*{APPENDIX A}  
\label{app1}

Here we provide the explicit form of Kraus operators for Markovian amplitude damping (AD) and phase damping (PD), and also the specific correlation functions corresponding to the case where the spin or CV components of our superposition state are affected by the environment.

\subsection{Amplitude damping} 

The general Kraus operators for a $d$-level system exposed to an amplitude damping (AD) channel characterized by the damping rate $\gamma$ are~\cite{Chuang}
\begin{equation}
\label{ADKO}
\hat{\cal A}^{AD}_{k}(t)=\sum_{n=k}^{d-1}\sqrt{
\begin{pmatrix}
n\\
k
\end{pmatrix}
}\,\eta^{(n-k)/2}(1-\eta)^{k/2}|n-k\rangle\langle n|
\end{equation}
with $k{=}0,..,d-1$, $\eta=e^{-\gamma{t}}$ and $[1-\eta]^{k/2}$ the probability that a Fock state $|n\rangle$ loses $k$ excitations within time $t$. 

\subsubsection{Spin damping}

For a spin-$1/2$ particle we have $d=2$ and the operator-sum decomposition consists of the following two elements
\begin{equation}
\hat{\cal A}^{AD}_0(t)=|0\rangle\langle0|+\sqrt{\eta}|1\rangle\langle1|,~~~\hat{\cal A}^{AD}_1(t)=\sqrt{1-\eta}|0\rangle\langle1|.
\end{equation}
We call ${\cal P}_{\rm AD}{=}\sqrt{1-\eta}$ the probability that the system loses one particle up to time $t$ and identify $\ket{0}$ ($\ket{1}$) with state $\ket{\downarrow}$ ($\ket{\uparrow}$). In this case, the correlation function 
for the Bell-CHSH non-locality test performed in the body of the manuscript takes the form (for simplicity, we assume $D,\beta\in\mathbb{R}$)
\begin{equation}
{\cal C}^{AD}_{spin}(\beta,\theta,\eta){=}e^{-2(\beta+D)^2}\frac{2\eta{-}1{-}e^{8\beta D}}{2}\cos\theta+\sqrt{\eta}e^{-2\beta^2}\sin\theta,
\end{equation}
which is such that $|{\cal C}^{AD}_{spin}(\beta,\theta,\eta)|{\le}1$.

\subsubsection{CV damping}

When considering a CV system, the summation in Eq.~\eqref{ADKO} extends to $d\rightarrow\infty$. Assume that the system is prepared in a coherent state $\ket{\xi}$. Such state is changed by the action of the $k^{\rm th}$ Kraus operator as
\begin{equation}
\hat{\cal A}^{AD}_{k}(t)|\xi\rangle=e^{-\frac{(1-\eta)|\xi|^2}{2}}\frac{(\xi\sqrt{1-\eta})^{k}}{\sqrt{k!}}|\sqrt{\eta}\xi\rangle,
\end{equation}
showing that a coherent state remains such under the action of an amplitude damping channel, although its amplitude is reduced. The Bell-CHSH correlation function in this case reads
\begin{equation}
\begin{split}
{\cal C}^{AD}_{cv}(\beta,\theta,\eta)&{=}\frac{1}{2}e^{-2(\beta+D\sqrt{\eta})^2}(1-e^{8\beta D\sqrt{\eta}})\cos\theta\\
&+e^{-2[\beta^2+{D^2}(1-\eta)]}\sin\theta.
\end{split}
\end{equation}

\subsection{Phase damping} 
The general Kraus operators for a $d$-level system exposed to a phase damping (PD) channel characterized by the rate $\mu$ are~\cite{liu}
\begin{equation}
\hat{\cal A}^{PD}_{k}(t)\equiv \hat{\cal A}^{PD}_{k}(\tau_{pd})=\sum_{n=0}^{d-1}e^{-\frac{1}{2}n^2\tau^2_{pd}}{\frac{(n\tau_{pd})^{k}}{\sqrt {k!}}}|n\rangle\langle n|
\end{equation}
where $\tau_{pd}=\mu t$ is the rescaled interaction time. 

\subsubsection{Spin damping}

For a spin-$1/2$ particle, the Kraus operators are
\begin{equation}
\begin{split}
\hat{\cal A}^{PD}_0(\tau_{pd})&{=}|0\rangle\langle0|+e^{-\frac{1}{2}\tau^2_{pd}}|1\rangle\langle1|,\\
\hat{\cal A}^{PD}_1(\tau_{pd}
)&{=}\sqrt{1-e^{-\tau^2_{pd}}}|1\rangle\langle1|,
\end{split}
\end{equation}
where ${\cal P}_{\rm PD}{=}\sqrt{1-e^{-\tau^2_{pd}}}$ is the probability that one excitation from the system is scattered by the environment.
The Bell-CHSH correlation function in this case reads
\begin{equation}
{\cal C}^{PD}_{spin}(\beta,\theta,\tau_{pd}){=}\frac {e^{-2(\beta+D)^2}}{2}(1-e^{8\beta D})\cos\theta{+}e^{-2\beta^2-\frac{\tau^2_{pd}}{2}}\sin\theta.
\end{equation}

\subsubsection{CV damping}

Again, let us assume that the CV system is prepared in a coherent state $\ket{\xi}$. Such state is changed by the action of the $k^{\rm th}$ Kraus operator as
\begin{equation}
\hat{\cal A}^{PD}_{k}(\tau_{pd})|\xi\rangle=\frac{\tau_{pd}^{k}}{\sqrt{k!}}\sum_{n=0}^{\infty}e^{-\frac{1}{2}(n^2\tau_{pd}^2+|\xi|^2)}\frac{n^{k}\xi^n}{\sqrt{n!}}|n\rangle.
\end{equation}
In the spin-$1/2$ basis $\{{|}{\uparrow}\rangle_{s},{|}{\downarrow}\rangle_s\}$, the Schr\"odigner cat state that has been affected only by a CV PD channel has the representation 
\begin{equation}
\rho^{PD}_{cv}=\frac{1}{2}
\left[\begin{matrix}
N_{++}(\tau_{pd}) & N_{+-}(\tau_{pd}) \\
N_{-+}(\tau_{pd}) & N_{--}(\tau_{pd})
\end{matrix}\right],
\end{equation}
with 
\begin{equation}
\begin{split}
N_{ij}(\tau_{pd})&=\sum_{k=0}^{\infty}\frac{\tau_{pd}^{2k}}{k!}\sum_{n,m=0}^{\infty}e^{-\frac{1}{2}(n^2\tau_{pd}^2+m^2\tau_{pd}^2+2|D|^2)}\\
&\times\frac{(nm)^{k}(i{D})^n(j{D^*}{})^m}{\sqrt{n!m!}}|n\rangle\langle m|,~~~~ (i,j=\pm).
\end{split}
\end{equation}
Let us first calculate the expectation value of the projection operator $\hat{\sigma}(\theta)$ defined in the body of the manuscript over the state $\rho^{PD}_{cv}$. We have
\begin{equation}
\Sigma(\theta,\tau_{pd})={\rm Tr}[\hat{\sigma}(\theta)\rho^{PD}_{cv}]=\frac{1}{2}\sum_{k=0}^{\infty}\frac{\tau_{pd}^{2k}}{k!}\sum_{n,m=0}^{\infty}\Omega |n\rangle\langle m|
\end{equation}
with
\begin{equation}
\begin{split} 
\Omega&=e^{-\frac{1}{2}(n^2\tau_{pd}^2+m^2\tau_{pd}^2+2|D|^2)}\frac{(nm)^{k}D^n(D^*)^m}{\sqrt{n!m!}}\\
&\times\{\sin\theta[(-1)^m+(-1)^n]+\cos\theta[(-1)^{(n+m)}-1]\}.
\end{split}
\end{equation}
The Bell-CHSH correlation function is then found by calculating the expectation value of the displaced parity operator $\hat{\Pi}(\beta)$ as
\begin{equation}
{\cal C}^{PD}_{cv}(\beta,\theta,\tau_{pd})={\rm Tr}[\hat{\Pi}(\beta)\Sigma(\theta,\tau_{pd})].
\end{equation}
During this calculation, we need to evaluate the expectation value of the parity operator $(-1)^{\hat n}{=}\sum_{n=0}^{\infty}(|2n{}\rangle\langle2n{}|{-}|2n{}{+}1\rangle\langle2n{}{+}1|)$ over displaced Fock states as \begin{equation}
\label{mddn}
\langle m|\hat{D}(\beta)(-1)^{\hat{n}}\hat{D}^{\dagger}(\beta)|n\rangle,
\end{equation}
which can be readily evaluated using~\cite{mauro}
\begin{equation}
\langle s|\hat{D}(\beta)|r\rangle=\sqrt{\frac{r!}{s!}}(\beta)^{s-r}e^{-\frac{|\beta|^2}{2}}L_r^{(s-r)}(|\beta|^2),~~~(s\ge r)
\end{equation}
where $L_p^{(l)}(x)$ is an associated Laguerre polynomial. We now take four different cases
\begin{enumerate}
\item  For $m\ge2n^{'}{+}1$ with $2n^{'}\ge n$. In this case we introduce the function
\begin{equation}
\begin{split}
&S_1=\sqrt{\frac{n!}{m!}}(\beta)^{m-n}e^{-|\beta|^2}(-1)^{2n'{-}n}[L_{2n'}^{(m{-}2n')}(|\beta|^2)\\
&\times L_n^{(2n'{-}n)}(|\beta|^2)+L_{2n'{+}1}^{(m{-}2n'{-}1)}(|\beta|^2)L_n^{(2n'{+}1{-}n)}(|\beta|^2)].
\end{split}
\end{equation}
\item For $2n^{'}\ge (m, n)$, we define
\begin{equation}
\begin{split}
&S_2=\frac{\sqrt{m!n!}}{2n'!}(-\beta^*)^{2n'-m}(-\beta)^{2n'-n}e^{-|\beta|^2}[L_m^{(2n'-m)}(|\beta|^2)\\
&\times L_n^{(2n'{-}n)}(|\beta|^2){-}\frac{|\beta|^2}{2n'+1}L_m^{(2n'{+}1{-}m)}(|\beta|^2)L_n^{(2n'{+}1{-}n)}(|\beta|^2)].
\end{split}
\end{equation}
\item For $(m, n)\ge2n^{'}{+}1$ we call
\begin{equation}
\begin{split}
&S_3=\frac{2n'!}{\sqrt{m!n!}}(\beta)^{m-2n'}(\beta^*)^{n-2n'}e^{-|\beta|^2}[L_{2n'}^{(m-2n')}(|\beta|^2)\\
&{\times}L_{2n'}^{(n-2n')}(|\beta|^2){-}\frac{2n'+1}{|\beta|^2}L_{2n'+1}^{(m-2n'-1)}(|\beta|^2)L_{2n'+1}^{(n-2n'-1)}(|\beta|^2)].
\end{split}
\end{equation}
\item For $n\ge2n^{'}{+}1$ and $2n^{'}\ge m$, we introduce               
\begin{equation}
\begin{split}
&S_4=\sqrt{\frac{m!}{n!}}(\beta^*)^{n-m}e^{-|\beta|^2}(-1)^{2n'{-}m}[L_m^{(2n'-m)}(|\beta|^2)\\
&\times L_{2n'}^{(n-2n')}(|\beta|^2)+L_m^{(2n'+1-m)}(|\beta|^2)L_{2n'+1}^{(n-2n'-1)}(|\beta|^2)].
\end{split}
\end{equation}
\end{enumerate}
With such definitions, the Bell-CHSH correlation function is finally obtained as 
\begin{equation}
\begin{split}
&C^{PD}_{cv}(\beta,\theta,\tau_{pd})=\\
&\frac{1}{2}\sum_{k=0}^{\infty}\frac{\tau^{2k}_{pd}}{k!}\sum_{n'=0}^{\infty}\left\{\sum_{m=2n'+1}^{\infty}\sum_{n=0}^{2n'}\Omega S_1+\sum_{m=0}^{2n'}\sum_{n=0}^{2n'}\Omega S_2\right.\\
&\left.+\sum_{m=2n'+1}^{\infty}\sum_{n=2n'+1}^{\infty}\Omega S_3+\sum_{n=2n'+1}^{\infty}\sum_{m=0}^{2n'}\Omega S_4\right\}.                  
\end{split}
\end{equation}

\renewcommand{\theequation}{B-\arabic{equation}}
\setcounter{equation}{0}
\section*{APPENDIX B}  

For the sake of providing a self-contained presentation, here we sketch a strategy to gather an analytic solution of the post-markovian master equation (ME) assessed in the body of the manuscript~\cite{SL}. We follow the lines sketched in Refs.~\cite{Briegel,Man} and first introduce the operator eigenbasis $\{\hat Q_k\}$ of the superoperator 
\begin{equation}
\begin{aligned}
\hat L(\rho_{s})&=\gamma_0(\nbar+1)(\hat\sigma_{-}\rho_{s}\hat\sigma_+-\frac{1}{2}\{\hat\sigma_+\hat\sigma_-,\rho_{s}\})\\
&+\gamma_0\nbar(\hat\sigma_+\rho_{s}\hat\sigma_--\frac{1}{2}\{\hat\sigma_-\hat\sigma_+,\rho_{s}\})
\end{aligned}
\end{equation}
 with $\hat\sigma_{\pm}=(\hat\sigma_x\pm i\hat\sigma_y)/2$ and $\rho_{s}$ the general density matrix of the spin system. The elements of such eigenbasis, which is commonly referred to as the {\it damping basis}, are such that
\begin{equation}
\label{LQ}
\hat L\hat Q_k=\lambda_k\hat Q_k,
\end{equation}
with $\{\lambda_k\}$ the set of corresponding eigenvalues. It is straightforward to find that~\cite{Briegel} 
\begin{equation}
\hat Q_1=\frac{1}{2}\left[\hat\openone-\frac{\hat\sigma_z}{(2\nbar+1)}\right], \hat Q_2=\hat\sigma_z, \hat Q_3=\hat\sigma_+, \hat Q_4=\hat\sigma_-
\end{equation}
with $\{\lambda_1=0, \lambda_2=2\lambda_{3,4}=-2\gamma_0(\nbar{+}1/2)\}$. In the damping basis, the density matrix of the system can be written as
\begin{equation}
\label{rhoQ}
\rho_s(t)=\sum_{k=1}^4 \alpha_k(t)\hat Q_k
\end{equation}
Substituting Eqs.~\eqref{LQ} and~\eqref{rhoQ} into the ME (10) in the main manuscript, the equation changes to 
\begin{equation}
\label{dopo}
\sum_l[\dot{\alpha}_l(t)-\beta_l(t)]\hat Q_l=0.
\end{equation}
where $\beta_k(t)=\int_0^{t}k(t')\lambda_k e^{\lambda_kt'}\alpha_k(t-t')dt'$. From this, we easily extract the set of integro-differential equations
\begin{equation}
\label{appb} 
\dot{\alpha}_l(t)=\beta_l(t)~~(l{=}1,..,4),
\end{equation}
which can be solved by resorting to Laplace transforms to get
\begin{equation} 
\alpha_p(t)={\rm Lap}^{-1}\left[\frac{1}{s-\lambda_p\tilde{k}(s-\lambda_p)}\right] \alpha_p(0),
\end{equation}
where $\tilde{X}(s){:}{=}{\rm Lap}[X(t)]$ is the Laplace transform of $X(t)$ and ${\rm Lap}^{-1}$ stands for the inverse transform. 
The figures of merit studied in the main paper are then easily determined from such solutions.


\begin{thebibliography}{99}
\bibitem{commentosize} This is understood by considering that $|\cos(2\tau_s)|^{N_s}{\simeq}e^{-2N_s\tau^2_s}$ for $N_s{\gg}1$ and $\tau_s\ll{1}$.
\end{thebibliography}
\end{document}